\documentclass[11pt]{article}
\usepackage{moriond,epsfig}

\def\Journal#1#2#3#4{{#1} {\bf #2}, #3 (#4)}

\def\NPB{{\em Nucl. Phys.} B}
\def\PLB{{\em Phys. Lett.}  B}
\def\PRL{\em Phys. Rev. Lett.}
\def\PRD{{\em Phys. Rev.} D}

\def\be{\begin{equation}}
\def\ee{\end{equation}}
\def\bea{\begin{eqnarray}}
\def\eea{\end{eqnarray}}

\begin{document}
\begin{flushright}
  MPI-PhT/2002-18 \\
  May 2002
\end{flushright}
\vspace*{4cm}
\title{TOP DECAY AND BOTTOM FRAGMENTATION IN NLO QCD
\footnote{Talk given by G. Corcella at XXXVIIth Rencontres de Moriond,
"QCD and Hadronic Interactions", 16-23 March 2002, Les Arcs, France.}
}

\author{$\underline{\mathrm{G. CORCELLA}}^1$ and A.D. MITOV$^2$}

\address{$^1$Max-Planck-Institut f\"ur Physik, Werner-Heisenberg-Institut,
D-80805 M\"unchen, Germany\\
$^2$ Department of Physics and Astronomy,
University of Rochester, Rochester, NY 14627, U.\ S.\ A.}

\maketitle\abstracts{We apply the method of perturbative fragmentation
to study bottom fragmentation in top decay. We present
the energy spectrum of $b$-quarks and $b$-flavoured hadrons in top decay.}

A reliable understanding of bottom-quark fragmentation in top-quark 
decay ($t\to bW$) will be fundamental to accurately measure the top 
properties, such as its mass, at present and future high-energy colliders.
At the LHC, final states with leptons and $J/\psi$, 
where the leptons are $W$-decay products and
the $J/\psi$ comes from the decay of a $b$-flavoured hadron, 
will be a promising channel to reconstruct the top mass $m_t$ \cite{avto}.
The error is estimated to be $\Delta m_t\simeq 1$~GeV
and the $b$-fragmentation is
the largest source of uncertainty, accounting for about 0.6 GeV. 
In this paper we investigate bottom fragmentation in top decay within 
the framework of perturbative fragmentation \cite{mele} 
and use phenomenological
models to describe the $b$-quark hadronization.

According to the factorization theorem,
the rate for the production of a $b$-hadron in top decay can be written,
up to power corrections, as
the following convolution:
\begin{equation} 
{1\over {\Gamma_0}} {{d\Gamma}\over {dx_B}}(x_B,m_t,m_b)=
{1\over {\Gamma_0}}{{d\hat\Gamma}\over {dx_b}}(x_b,m_t,\mu_F)
\otimes D_p (x_b,m_b,\mu_F)\otimes D_{np}(x_B).
\label{fact}
\end{equation}
In Eq.~(\ref{fact}), $\Gamma_0$ is the Born width of the process $t\to bW$, 
$d\hat\Gamma/dx_b$ is the rate for the production of a massless
$b$ quark in top decay (coefficient function), 
$D_p (x_b,m_b,\mu_F)$ is the perturbative 
fragmentation function which 
expresses the transition from a massless $b$ into a
massive $b$ quark, $\mu_F$ is the factorization scale, 
$D_{np}$ is the non-perturbative fragmentation
function. $x_b$ and $x_B$ are the normalized $b$-quark and
$b$-hadron energy fractions in top decay \cite{cormit}.

We have computed the rate for the production of a massless $b$ quark in top
decay in dimensional regularization and subtracted the collinear singularity
in the $\overline{\mathrm{MS}}$ factorization scheme. 
We have got the following $\overline{\mathrm{MS}}$ coefficient function:
\begin{equation}
\left({1\over{\Gamma_0}}{{d\hat\Gamma_b}\over {dz}}\right)^{\overline
{\mathrm{MS}}}=
\delta(1-z)+{{\alpha_S(\mu)}\over{2\pi}}\hat A_1 (z),
\label{diff}
\end{equation}
\noindent 
with
\begin{eqnarray}
\hat A_1 (z) &=& C_F\left\{
\left[ {{1+z^2}\over{(1-z)_+}}+{3\over 2}\delta (1-z)\right]
\left[\log{{m_t^2}\over {\mu_F^2}}+2 {{1+w}\over {1+2w}}-2\log (1-w)\right]
\right.
\nonumber\\
&+&{{1+z^2}\over {(1-z)_+}} \left[ 4\log \left[(1-w)z\right]-
{1\over{1+2w}}\right]
-{{4z}\over{(1-z)_+}}\left[1-{{w(1-w)(1-z)^2}\over
{(1+2w)(1-(1-w)z)}}\right]\nonumber\\
&+&2(1+z^2)\left[\left({1\over {1-z}}\log(1-z) \right)_+-{1\over {1-z}}
\log z\right]\nonumber\\
&+& \delta(1-z) \left[4{\mathrm {Li}}_2 (1-w)+
2\log(1-w)\log w -{{2\pi^2}\over 3}
+{{1+8w}\over {1+2w}}\log (1-w)\right.\nonumber\\
&-&{2w\over{1-w}}\log w+
{{3w}\over {1+2w}}-9\Bigg]\Bigg\}.
\label{cms}
\end{eqnarray}
The value of the perturbative fragmentation function $D_p(x_b,m_b,\mu_F)$ 
at any scale $\mu_F$ can be obtained by solving the
Dokshitzer--Gribov--Lipatov--Altarelli--Parisi (DGLAP)
evolution equations, once an initial condition at a scale
$\mu_{0F}$ is given. 
The initial condition of the perturbative fragmentation function 
reads \cite{mele}:
\begin{equation}
D_p(x_b,\mu_{0F},m_b)=\delta(1-x_b)+{{\alpha_S(\mu_0)C_F}\over{2\pi}}
\left[{{1+x_b^2}\over{1-x_b}}\left(\log {{\mu_{0F}^2}\over{m_b^2}}-
2\log (1-x_b)-1\right)\right]_+.
\label{dbb}
\end{equation}
The result in Eq.~(\ref{dbb}) is process 
independent \cite{cc} and can hence be used in top decay as well.
Moreover, it has been shown \cite{cormit}, by comparing the rates for 
massless and massive $b$-quark production in top decay, 
that the coefficient function 
(\ref{cms}) is indeed consistent with Eq.~(\ref{dbb}).
Solving the DGLAP equations for the 
evolution $\mu_{0F}\to \mu_F$, with a NLO kernel,
allows one to resum logarithms $\log(\mu_F^2/\mu_{0F}^2)$ with NLL 
accuracy.
If we set $\mu_{0F}\simeq m_b$ and $\mu_F\simeq m_t$, we resum
large logarithms $\sim \log(m_t^2/m_b^2)$ which appear
in the massive, unevolved, fixed-order calculation of 
$d\Gamma/dx_b$ \cite{cormit}.
The coefficient function (\ref{cms}) and the initial condition
(\ref{dbb}) contain terms which get large once $x_b\to 1$. This  
corresponds
to soft-gluon radiation in top decay. Soft terms in the initial condition
of the perturbative fragmentation  function are process independent and
have been resummed with next-to-leading logarithmic accuracy \cite{cc}.
Soft-gluon contributions in the coefficient function are process dependent 
and their resummation is
currently in progress.

\begin{figure}
\begin{center}
\psfig{figure=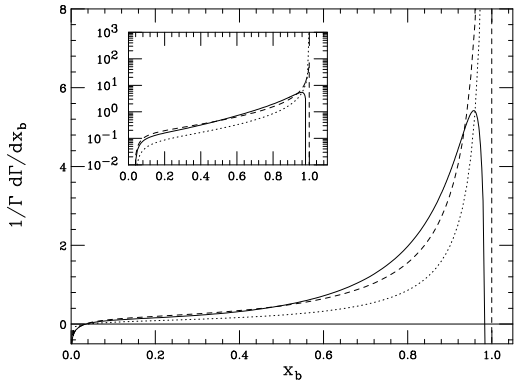,height=2.in}
\caption{$b$-quark spectrum in top decay according to the perturbative
fragmentation method, 
with (solid) and without (dashes) NLL
soft-gluon resummation in the initial condition (\ref{dbb}), and
according to the ${\cal O}(\alpha_S)$ massive result (dots).
In the inset figure, the same distributions are plotted on a logarithmic scale.
\label{figpart}}
\end{center}
\end{figure}

In Fig.~\ref{figpart} we present the $b$-quark energy
distribution in top decay according to the approach just
described. We note that the inclusion of soft-gluon resummation in
the initial condition of the perturbative fragmentation function smoothens
out the distribution and the $x_b$ spectrum exhibits the so-called 
Sudakov peak. Also shown is the massive ${\cal O}(\alpha_S)$ result,
which lies below the other two distributions throughout the spectrum and 
is divergent as $x_b\to 1$.
It has also been found that the implementation of soft resummation
yields a milder dependence of observables on the factorization scale
$\mu_{0F}$ which enters Eq.~(\ref{dbb}) \cite{cormit}.
\par
We would like to make predictions for the spectrum of $b$-flavoured hadrons in 
top decay. For this purpose we use some phenomenological models which
contain tunable parameters which are to be
fitted to experimental data.
We consider a power law with two parameters:
\begin{equation}
D_{np}(x;\alpha,\beta)={1\over{B(\beta +1,\alpha +1)}}(1-x)^\alpha x^\beta,
\label{ab}
\end{equation}
the model of Kartvelishvili et al. \cite{kart}
\begin{equation}
D_{np}(x;\delta)=(1+\delta)(2+\delta) (1-x) x^\delta
\label{kk}
\end{equation}           
and the non-perturbative fragmentation function of Peterson et 
al. \cite{peterson}:
\begin{equation}
D_{np}(x;\epsilon)={A\over {x[1-1/x-\epsilon/(1-x)]^2}}.
\label{peter}
\end{equation}
In Eq.~(\ref{ab}), $B(x,y)$ is the Euler Beta function;
in (\ref{peter}) $A$ is a normalization constant.
We tune such models to $e^+e^-$ data from the 
SLD \cite{sld} and ALEPH \cite{aleph} Collaborations. The SLD data 
refers to $b$-flavoured mesons and baryons and the ALEPH to mesons.
When we do the fits, we must describe the $e^+e^-\to b\bar b$ process 
within the same framework as 
for top decay, i.e. we use the perturbative fragmentation method,
NLL DGLAP evolution and NLL soft-gluon resummation
in the initial condition (\ref{dbb}).

\begin{figure}
\begin{center}
\psfig{figure=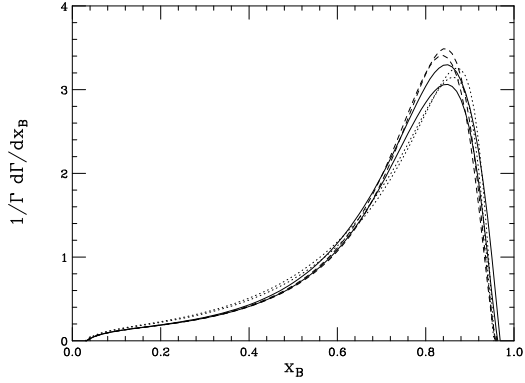,height=2.in}
\caption{$b$-hadron spectrum in top decay, describing
the hadronization according to the power law with two parameters 
(solid), the Kartvelishvili 
(dashes) and the Peterson (dots) models, fitted to the ALEPH data.
For any model, we plot the edges of a band corresponding to 
one-standard-deviation confidence level.
\label{fitaleph}}
\end{center}
\end{figure}

\begin{figure}
\begin{center}
\psfig{figure=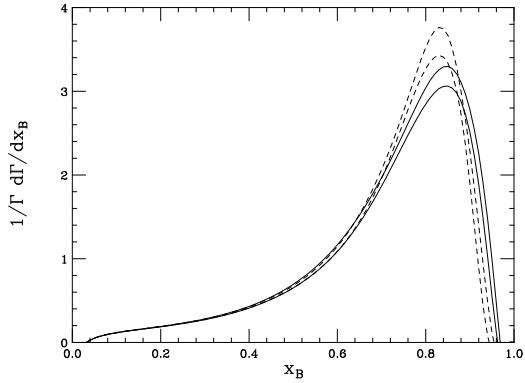,height=2.in}
\caption{As in Fig.~\ref{fitaleph}, using the power law, but fitted to
ALEPH (solid) and SLD (dashes).
\label{fitsld}}
\end{center}
\end{figure}

In Table~\ref{fit} we report on the parameters which yield the best fits of
the hadronization models, along with the corresponding standard deviations 
and the $\chi^2$ per degree of freedom.
\begin{table}[t]
\caption{Best-fit values for the parameters contained 
in the non perturbative fragmentation functions.\label{fit}}
\vspace{0.4cm}
\begin{center}
\begin{tabular}{||l|r|r||}\hline
&ALEPH\hspace{1.05cm} &SLD\hspace{1.45cm} \\ \hline 
\hspace{1.cm}$\alpha$&$0.31\pm 0.15$\hspace{0.79cm} &$1.88\pm 0.42$
\hspace{0.78cm} \\ \hline
\hspace{1.cm}$\beta$&$13.21\pm 1.62$\hspace{0.8cm} 
&$27.04\pm 4.02$\hspace{0.92cm} \\ \hline
$\chi^2(\alpha,\beta)$/dof&2.62/14\hspace{1.2cm} 
&11.12/16\hspace{1.2cm} \\ \hline
\hspace{1.cm}$\delta$&$20.39\pm 0.77$\hspace{0.8cm} 
&$18.80\pm 0.60$\hspace{0.92cm} \\ \hline
\hspace{0.2cm}$\chi^2(\delta)$/dof&17.27/15\hspace{1.2cm} 
&17.46/17\hspace{1.1cm}  \\ \hline
\hspace{1.cm}$\epsilon$&$(1.12\pm 0.16)\times 10^{-3}$
&$(1.17\pm 0.10)\times 10^{-3}$\\ \hline
\hspace{0.2cm}$\chi^2(\epsilon)$/dof&22.96/15\hspace{1.1cm}
&130.80/17\hspace{1.1cm} \\ \hline
\end{tabular}
\end{center}
\end{table}
We note that the model in Eq.~(\ref{ab}) fits best the data, although
the best-fit parameters show pretty-large uncertainties.
The Kartvelishvili model well reproduces both ALEPH and SLD data, while
the Peterson model marginally agrees with the ALEPH data and is
inconsistent with the SLD one. Moreover, the values of
the parameters $\delta$ and $\epsilon$, fitted to ALEPH and SLD, are
in agreement within two standard deviations.
In Fig.~\ref{fitaleph} we show our prediction for the $b$-hadron spectrum in
top decay, using all three hadronization models, fitted to ALEPH.
We note that each model yields a statistically-different prediction for
the $x_B$ spectrum, the Peterson-based distribution being peaked at
larger $x_B$ values. In Fig.~\ref{fitsld} we compare the predictions
obtained using the power law with two parameters, but fitted to 
ALEPH and SLD. We observe that the spectra are distinguishable; such
a difference may be related to the different hadron types that the
two experiments have reconstructed.

In summary, we have reviewed the main results obtained once we studied the
fragmentation of $b$ quark in top decay using the perturbative
fragmentation function approach and presented the energy distribution of
$b$ quarks and $b$-flavoured hadrons in top decay.

\end{document}